\documentclass[11pt,a4paper,twocolumn]{book}
\usepackage{jnpcs}
\usepackage{graphics} %   <------- for LatTeX + PS
\usepackage{epsfig}
\newcommand{\beq}{\begin{equation}}
\newcommand{\eeq}{\end{equation}}
\newcommand{\beqa}{\begin{eqnarray}}
\newcommand{\eeqa}{\end{eqnarray}}
\newcommand{\ba}{\begin{array}}
\newcommand{\ea}{\end{array}} 

\ntitle{Colored $1/f^{\alpha}$ noise and the Order to Chaos 
Transition \\in Quantum Mechanics}%<------title

\nauthor{L. Salasnich}%  <--------- Authors

\naddress{CNR-INFM, Research Unit of Milano University, \\
Dipartimento di Fisica, Universit\`a di Milano, \\
      \it   via Celoria 16, 20133 Milano, Italy \\
{\small \rm
   E-mail: salasnich@.mi.infm.it } \\ 
}%<------address & e-mail

\ndata{
(Received 30 October 2005) }

\nabstract{ 
The spectral statistic $\delta_n$ measures the fluctuations of 
the number of energy levels around its mean value. 
It has been shown that chaotic quantum systems display $1/f$ noise 
(pink noise) in the power spectrum $S(f)$ of the $\delta_n$ statistic, 
whereas integrable ones exibit $1/f^2$ noise (brown noise). 
These results have been explained on the basis of the
random matrix theory and periodic orbit theory. 
Recently we have analyzed the order to chaos 
transition in terms of the power spectrum $S(f)$ by using the Robnik 
billiard (Phys. Rev. Lett. {\bf 94}, 084101 (2005)). 
We have numerically found a net power law $1/f^{\alpha}$, 
with $1\leq \alpha \leq 2$, at all the transition stages. 
Similar results have been obtained by 
Santhanam and Bandyopadhyay (Phys. Rev. Lett. {\bf 95}, 114101 (2005)) 
analyzing two coupled quartic oscillators 
and a quantum kicked top. All these numerical results suggest 
that the exponent $\alpha$ is related to the chaotic 
component of the classical phase space of the quantum billiard, 
but a satisfactory theoretical explanation is still lacking.}

\nkeywords{Quantum chaos; Numerical simulations of chaotic systems}

\nPACSnumbers{05.55.Mt; 05.45.Pq}
\begin{document}
\DeclareGraphicsExtensions{.jpg,.pdf,.mps,.png} %   <------- for PDFLatTeX
\firstpage{1} \nlpage{2} \nvolume{4} \nnumber{1} \nyear{2006}
\def\nfpage{\thepage}
\thispagestyle{myheadings} \npcstitle
%*****************   The Body of the Article:   *************************

Quantum chaos is the study of quantum systems which are 
classically chaotic \cite{r1}. Many experiments and numerical 
results support a strong relationship between the energy level fluctuation 
properties of a quantum systems and the dynamical behavior of 
its classical analog \cite{r1}. 
In the past years many energy-level statistics have been proposed 
to characterize the spectral fluctuations. Recently, 
Relano {\it et al.} \cite{r2} have introduced 
the energy-level statistic $\delta_n$, 
showing that chaotic quantum systems display $1/f$ pink noise 
in the power spectrum $S(f)$ of the $\delta_n$ statistic, 
whereas integrable ones exibit $1/f^2$ brown noise \cite{r2}. 
An open problem is the behavior of the power spectrum 
$S(f)$ for systems in the mixed regime between order and chaos. 
In this brief memoir we discuss very recent analytic 
and numerical results on $S(f)$ at the transition stages 
between order and chaos. 
\par 
Let us consider a quantum system whose Hamiltonian $\hat H$ has 
a discrete set of energy levels $E_i$. 
The staircase function $N(E)$, which gives the number of energy levels 
up to the energy $E$, can be written as 
$$ 
N(E) = {\bar N}(E) + N_{osc}(E) \; , 
$$ 
where ${\bar N}(E)=E$ is the averaged number of levels 
and $N_{osc}(E)$ is the fluctuating part. $N_{osc}(E)$ is supposed 
to belong to a universality class, which should only depend on 
the integrability or chaoticity of the classical analog \cite{r1}.  
The unfolded energy levels are given by 
$
\epsilon_i = {\bar N}(E_i) 
$, 
such that the energy is measured in units 
of the mean level speacing.  
The spectral statistic $\delta_n$ is defined as 
$$ 
\delta_n = 
\sum_{i=1}^n\left(s_i-\left<s\right>\right)=\epsilon_{n+1}-\epsilon_1-n,
$$ 
where $s_i=\epsilon_{i+1}-\epsilon_i$, and $<s>=1$ is the average value 
of $s_i$. 
Thus $\delta_n$ represents the fluctuation of the $n$-th excited state 
with respect to its mean value.  
Formally $\delta_n$ is similar to a time series where the
level order index $n$ plays the role of a discrete time. 
Therefore the statistical behavior of level fluctuations can be investigated
studying the power spectrum $S(k)$ of the signal, given by
$$
S(k)=\left| {1\over \sqrt{M}} \sum_{n=1}^M \delta_n \exp{\left(
{-2\pi i k n\over M} \right)} \right|^2 \; , 
$$ 
where $M$ is the size of the series and $f=2{\pi} k/M$ plays the role
of a frequency. 
\par
The Madrid group \cite{r3} and independently Robnik \cite{r4} 
have analytically proved that, under the conditions 
$M\gg 1$ and $k/M\ll 1$, the averaged power spectrum of the 
energy-level statistic $\delta_n$ is given by 
$$
\langle S(k) \rangle 
= \left\{  \ba{cc} 
{2 \over \beta} {M\over k} & \mbox{for chaotic systems} \\ \\
{M^2\over k^2} & \mbox{for integrable systems} \\
\ea 
\right. \; , 
$$ 
where $\beta$ depends on the symmetry of the Gaussian ensemble: 
$\beta=1$ for the Gaussian Ortogonal Ensemble (GOE), $\beta=2$ for the 
Gaussian Unitary Ensemble (GUE), and $\beta=4$ for the Gaussian Symplectic 
Ensemble (GSE). 
\par 
An interesting and important question is: what is 
the behavior of $\langle S_M(k)\rangle$ 
for systems in the mixed regime between order and chaos?  
On the basis of the principle of uniform semiclassical condensation 
(PUSC), Robnik has derived, under the conditions $M\gg 1$ and 
$k/(M\mu_C)\ll 1$, the formula 
$$
\langle S(k) \rangle = {M^2\over k^2} 
\left( \mu_R + {k\over M} \right) \; , 
$$ 
where $\mu_R$ and $\mu_C$ are the regular and chaotic 
fractions of the energy spectrum, 
such that $\mu_R + \mu_C =1$ \cite{r4}. 
To verify this prediction, we have recently 
analyzed \cite{r5} the Pascal's snail, also called 
as Robnik billiard \cite{r6}. 
The boundary of the Robnik billiard is defined as the set of points
$w$ in the complex plane ${\bf C}$ which satisfy the equation
$
w=z+\lambda z^2  
$, 
where $|z|=1$ and $\lambda$ is the deformation 
parameter. This billiard exhibits a smooth transition 
from the integrable case ($\lambda =0$)
to an almost chaotic case ($1/4\leq \lambda \leq 1/2$) \cite{r6}.  
The quantum energy levels $E_n$ of the Robnik billiard 
have been numerically calculated by solving the stationary 
Schr\"odinger equation of a free 
particle whose wave function $\psi(w)$ is zero at the boundary of the
billiard. As shown in Fig. 1, we have numerically found a net power law 
$$
\langle S(k) \rangle \sim {1\over k^{\alpha}} \; , 
$$  
with $1\le \alpha \le 2$, at all the transition stages, 
in contrast with the anaytical prediction of Robnik,
based on the PUSC. In Fig. 2 we plot $\alpha -1$ as a 
function of the deformation parameter $\lambda$ of the Robnik 
billiard. The figure shows that $\alpha -1$ 
has a behavior close to that of the regular component 
$\rho_1^{cl}$ of the classical phase space and also 
to that of $1-\omega$, where $\omega$ is the Brody 
parameter that measures the chaoticity of the 
energy spectrum \cite{r1}.  

\begin{figure}[ht!]% Fig.1.
     \leavevmode
\centering
\includegraphics[width=2.9in, height=3.5in, angle=-90]{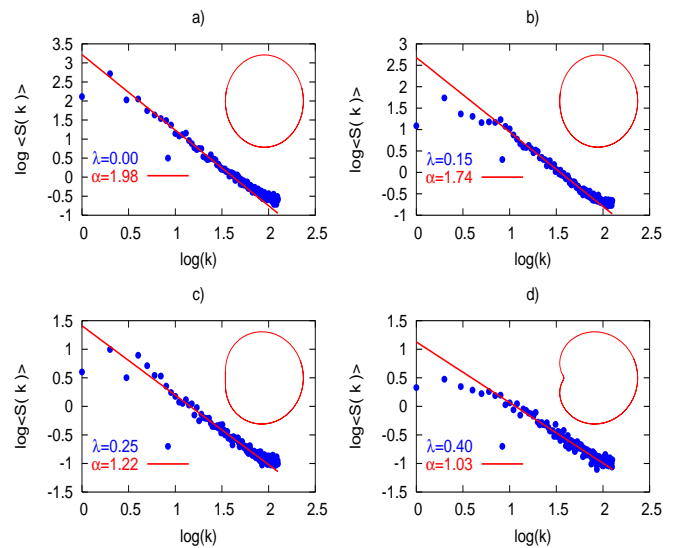}
\caption{Average power spectrum $\langle S(k) \rangle$ of the 
statistic $\delta_n$ for the odd parity energy levels corresponding to
the shapes of the Robnik billiard inserted in the figures. 
The solid line is the best fit to the power law $1/k^{\alpha}$.} 
\end{figure}

\begin{figure}[ht!]% Fig.2.
     \leavevmode
\centering
\includegraphics[width=2.8in, height=2.5in, angle=0]{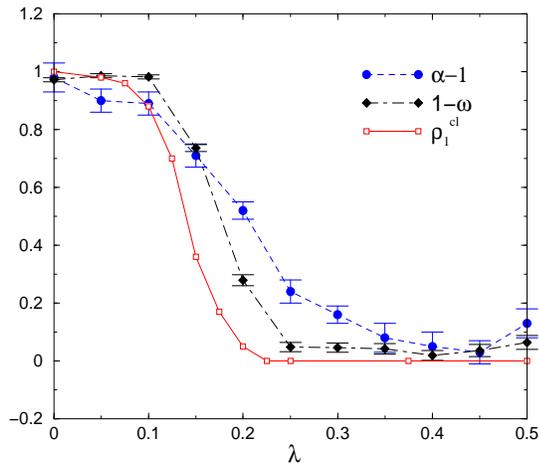}
\caption{Behavior of the power spectrum exponent $\alpha$, 
the Brody parameter $\omega$, and the 
fraction $\rho_1^{cl}$ of regular classical trajectories 
of the Robnik billiard as functions of the deformation 
parameter $\lambda$. } 
\end{figure}

It is important to observe that the $\delta_n$ statistic 
has been investigated also by Santhanam and 
Bandyopadhyay \cite{r7}. 
They have analyzed two quantum systems which have 
a classical transition from order to chaos: 
the two coupled quartic oscillators and the kicked top.  
\par 
The Hamiltonian of the two coupled quartic oscillators 
is given by 
$$
H = p^{2}_{1} + p^{2}_{2} + q_1^{4}+q_2^{4} 
+ g \; q_1^2 q_2^2 
\; , 
$$
where $g$ is the strength of the nonlinear coupling between 
the two quartic oscillators. The variables $q_i$ ($i=1,2$) 
are the coordinates of the two oscillators and $p_i$ are 
the conjugate momenta. As discussed in \cite{r7}, 
this system is classically integrable for $g=0$ and
the phase space is predominantly chaotic for $g>6$. 
In Fig. 3 we show the power spectrum $S(\tau )$ of $\delta_n$ 
calculated by Santhanam and Bandyopadhyay for increasing 
values of the nonlinear strength $g$ \cite{r7}. 
As in the case of the Robnik billiard 
\cite{r6}, also for the two coupled quartic oscillators there is 
numerical evidence of a net power law in the average 
power spectrum $\langle S(\tau) \rangle $. 
Similar results have been obtained by Santhanam and Bandyopadhyay 
analyzing the kicked top \cite{r7}. 

\begin{figure}[ht!]% Fig.3.
     \leavevmode
\centering
\includegraphics[width=3.in, height=2.5in, angle=0]{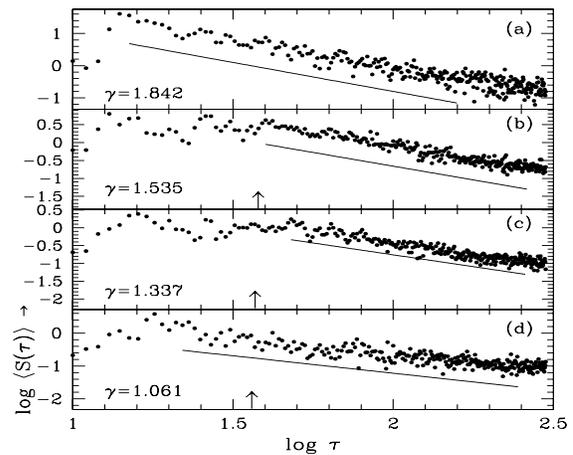}
\caption{Power spectrum $S(\tau )$ of the $\delta_n$ statistic 
for the two coupled quartic oscillators with increasing 
values of the nonlinear strength: $g=0$ (a), 
$g=7.5$ (b), $g=11.5$ (c), and $g =19.5$ (d). 
The solid lines are the least squares fit 
with intercept shifted for clarity. 
The slope $\gamma$ is indicated in each graph. 
Adapted from [7].} 
\end{figure}

In conclusion, while for a fully integrable and a fully chaotic 
system the colored noise of $\delta_n$ has been theoretically 
explained, for mixed systems an explanation 
of the power-law behavior $S(k)\sim 1/k^\alpha$ is still lacking. 
Some important questions need to be answered: 
Is the exponent $\alpha$ of the power law a measure of 
the chaoticity of the classical phase space? 
Does the behavior of $S(k)$ change in the semiclassical limit 
of very high level density, as predicted by Robnik using the Principle 
of Uniform Semiclassical Condensation? 
A critical discussion of colored noise in quantum chaos and 
the derivation of new spectral statistics, whose power spectrum 
is characterized by colored noise, can be found in \cite{r8}.

\end{document}